\documentclass[letterpaper]{article} % DO NOT CHANGE THIS
\usepackage{aaai25}  % DO NOT CHANGE THIS
\usepackage{times}  % DO NOT CHANGE THIS
\usepackage{helvet}  % DO NOT CHANGE THIS
\usepackage{courier}  % DO NOT CHANGE THIS
\usepackage[hyphens]{url}  % DO NOT CHANGE THIS
\usepackage{graphicx} % DO NOT CHANGE THIS
\usepackage{natbib}  % DO NOT CHANGE THIS AND DO NOT ADD ANY OPTIONS TO IT
\usepackage{caption} % DO NOT CHANGE THIS AND DO NOT ADD ANY OPTIONS TO IT
\usepackage{pdfpages}
\usepackage{amsmath}
\usepackage{algorithm}
\usepackage{algpseudocode}
\usepackage{amsthm}
\usepackage{xcolor}
\usepackage{tikz}
\usepackage{newfloat}
\usepackage{listings}
\usepackage{xcolor}
\usepackage [english]{babel}
\usepackage [autostyle, english = american]{csquotes}
\usepackage{enumitem}

\usepackage{xcolor}
\usepackage{textcomp}

\MakeOuterQuote{"}
\DeclareCaptionStyle{ruled}{labelfont=normalfont,labelsep=colon,strut=off} % DO NOT CHANGE THIS
\floatstyle{ruled}
\newfloat{listing}{tb}{lst}{}
\floatname{listing}{Listing}
\title{Enhancing Joint Human-AI Inference in Robot Missions:\\ A  Confidence-Based Approach}
\author{
    Duc-An Nguyen\textsuperscript{\rm 1}\thanks{Corresponding author: annguyen@robots.ox.ac.uk},
    Clara Colombatto\textsuperscript{\rm 2},
    Steve Fleming\textsuperscript{\rm 3},
    Ingmar Posner\textsuperscript{\rm 1},\\
    Nick Hawes\textsuperscript{\rm 1},
    Raunak Bhattacharyya\textsuperscript{\rm 4}
}
\affiliations{
    \textsuperscript{\rm 1}Oxford Robotics Institute, University of Oxford\\
    \textsuperscript{\rm 2}Department of Psychology, University of Waterloo\\
    \textsuperscript{\rm 3}Department of Experimental Psychology, University College London\\
    \textsuperscript{\rm 4}Yardi School of Artificial Intelligence, Indian Institute of Technology Delhi\\
}

\usepackage{bibentry}
\begin{document}

\maketitle

\begin{abstract}
Joint human-AI inference holds immense potential to improve outcomes in human-supervised robot missions.
Current day missions are generally in the AI-assisted setting, where the human operator makes the final inference based on the AI recommendation.
However, due to failures in human judgement on when to accept or reject the AI recommendation, \textit{complementarity} is rarely achieved.
We investigate joint human-AI inference where the inference made with higher confidence is selected.
Through a user study with $N=100$ participants on a representative simulated robot teleoperation task, specifically studying the inference of robots' control delays we show that: a) Joint inference accuracy is higher and its extent is regulated by the confidence calibration of the AI agent, and b) Humans change their inferences based on AI recommendations and the extent and direction of this change is also regulated by the confidence calibration of the AI agent.
Interestingly, our results show that pairing poorly-calibrated AI-DSS with humans hurts performance instead of helping the team, reiterating the need for AI-based decision support systems with good metacognitive sensitivity. 
To the best of our knowledge, our study presents the first application of a maximum-confidence-based heuristic for joint human-AI inference within a simulated robot teleoperation task.

\end{abstract}


% \input{main-sections/introduction}
% \input{main-sections/related-work}
% \input{main-sections/method}
% \input{main-sections/human-experiment}
% \input{main-sections/results}
% \input{main-sections/discussion}
% % \section{Conclusion}
% % \vspace{40pt}
% \input{main-sections/ethical-statement}

\section{Introduction}

% \an{Note start:}
% \begin{enumerate}
%     \item ECAI format
%     \item Fig.5; Fig.8 poorly: red, and more red. Good: green and more green.
%     \item Fig.7, change color code too, and make a version of joined into 1 fig. consistent font size
%     \item Fig.6 to section B
%     \item fig. 5 aggregate hard and easy into 1.
%     \item Methodology: C. 2 group of human pairing with 2 calibrated systems. The participant randomly paried with 1 type of calibrated systems (Welll, or Poorly). we collect 50 participant interact with each type. then after exluding, we have: we paired the human decision maker with well-calibrated (AUROC2 ≥ 0.65) AI-DSS (42 participants) versus poorly-calibrated (AUROC2 ≤ 0.55) AI-DSS
%     \item histogram fig
%     \item interface fig, to be cut into 3 pieces, an look into Ming Yin Fig 2.
% \end{enumerate}

% \an{Note end:}

\textit{Inference} is a major challenge for human supervisors of robot missions.
For example, missions such as search and rescue, defense, firefighting, and space exploration~\cite{schoonderwoerd2022design,kase2022future,verhagen2024meaningful} require humans to make inferences under communication latencies, incomplete information, and time pressure.
AI-based decision support systems (henceforth, AI-DSS) hold immense potential to improve inference (and thereby, decision making) in human-supervised robot missions.

While AI-DSS have been used in robot missions such as urban search and rescue~\cite{hong2019investigating}, their use has largely been explored in \textit{AI-assisted} configurations, i.e., where the final decision to accept or reject the AI-DSS recommendation lies with the human operator. 
Similarly to non-robotic tasks such as recidivism risk prediction~\cite{dss_whentoadvice} and medical diagnosis~\cite{multiplehumans_ijcai22}, the AI-DSS is often used as a ``second opinion'' system~\cite{chong2022human, chiou2021mixed}.

However, humans often make incorrect inferences even when provided with recommendations by an AI-DSS~\cite{natarajan2025human}. 
Their failure to discern when to accept or reject AI input hinders team performance and \textit{human-AI complementarity} is rarely achieved~\cite{bansal2021does, peng2025no}.
For example, human decision makers may change from an initially correct inference to an incorrect one based on the AI-DSS's recommendation. 
% Unlike catastrophic hardware failures, robot control delays are subtle, continuous, and time-varying, stemming from factors like software performance or network fluctuations. They are not detectable from single sensor readings, revealing themselves only through sequential observations and active robot interaction. This subtlety makes them an insidious threat to mission performance and safety. Standard paradigms in robot teleoperation have fundamental limitations; purely algorithmic approaches require explicit fault models or large datasets often unavailable for novel delay sources, leading to poorly timed commands and failed actions. Conversely, human operators constantly monitoring minute timing differences face immense cognitive strain, causing performance degradation and error.
% This can result in the gradual degradation of overall decision-making performance, particularly in complex, fast-paced, and long-term robot missions. However, it is infeasible to ask humans for precise numerical 'This robot is running 47 milliseconds slower.' H Exceptionally, humans are skilled at providing qualitative, intuitive judgments such as 'That robot seems sluggish.' This inherent human capability, particularly in metacognitive assessment, suggests qualitative input like confidence levels is a powerful mechanism for human-AI collaboration. 
Therefore, joint human-AI decision making has been proposed as an alternative paradigm~\cite{rastogi2023taxonomy,zahedi2021human} where the final team inference can be based on either of the individual inferences made by the human and the AI-DSS.
However, whose inference should finally be chosen remains an open problem~\cite{natarajan2025human,methnani2024s}.

Research from the field of human-human joint decision making proposes an approach based on \textit{confidence calibration}~\cite{fleming2014measure}.
Specifically, \textit{confidence} has been shown as a viable heuristic for joint inference.
Known as Maximum Confidence Slating (MCS), this approach accepts the inference made with \textit{higher confidence}, assuming that each individual can monitor their own performance and can communicate their confidence accurately~\cite{bahrami2010optimally,koriat2012two}.
On simple inference tasks based on static, visual inputs, this approach has shown higher team accuracy than either individual alone.
This raises the prospect of using confidence estimates in joint human-AI inference for robot missions.

However, the challenge of inference in robot missions arises out of three factors: 1) latent variables, 2) temporal processes, and 3) inference based on actions.
Specifically, human supervisors have to make inferences about unobserved variables such as robot state~\cite{ramesh2023experimental}.
These inferences often have to be made on the basis of video, which is an unfolding temporal process as opposed to static, visual representations of information.
Finally, the human supervisor has to often take actions, such as operating the robot, to be able to gather information to make inferences.
This is in contrast to passive inference where the images are provided to the human.
To the best of our knowledge, while joint human-AI inference has been studied in tasks involving either latent variables~\cite{mahmood2024designing}, temporal processes~\cite{inkpen2023advancing}, or on the basis of actions, there is a gap in the literature on joint human-AI inference for tasks involving all three of these challenges.

In this paper, we investigate a representative simulated robotic task that poses these three challenges and allows us to systematically study their impact on both individual human and joint human-AI inference, respectively.
We conduct a behavioral study of $100$ participants where they have to select between two robots undergoing control delays.
Such tasks often arise in robotic disaster response where human operators have to select the appropriate robot to localise survivors~\cite{wang2023development}.
The participants have to infer which robot has lesser delay (a latent variable) by observing the robot's response through video (temporal process) while teleoperating the robot (taking actions).
The participants are paired with an AI-DSS which also provides its inference. 
% While this paper does not introduce a new AI model for continuous delay classification, we directly investigate leveraging human-AI confidence exchange on a Likert scale to enhance joint inference accuracy. 
The AI-DSS inference is based on human subject data gathered separately.
Both the human and the AI-DSS provide a confidence value associated to their inference and the joint inference is decided via MCS, i.e., inference made with higher confidence. 

We ask the following overarching research questions:
\begin{enumerate}[label=\Roman*), topsep=3pt]
    \item How is human inference impacted by the AI-DSS inference and its confidence characteristics?
    \item Does joint human-AI inference using MCS lead to better performance as compared to the AI-assisted inference setting?
\end{enumerate}

We assess the performance of MCS-based joint inference by comparing the resulting accuracy to that of individual human and AI-assisted human inference, respectively.
We investigate the impact of confidence alignment between humans and AI on the accuracy of joint inference. Confidence alignment refers to the distributional distance between the confidence levels of the human and the AI-DSS in the inferencing process.
In a between-subject study, we randomly vary whether participants are paired with a well-calibrated AI-DSS or poorly-calibrated AI-DSS.
We measure the impact of these experimental conditions on factors such as the participants' accuracy, whether they change their original inference in response to the AI-DSS suggestion, and the correctness of their changed inferences.
Further, we investigate how the relative confidence calibration between the human and AI-DSS impacts joint team inference performance.

This study is, to the best of our knowledge, the first to examine the MCS algorithm for human-AI joint inference in a representative robotic task which poses the challenges of latent variables, temporal processes and actions for inference.
Our study yields the following findings:

\begin{enumerate}[label=\Roman*),topsep = 2pt]
    % \item Humans show more propensity to change their initial inferences when the AI-DSS provides its inferences with higher confidence and when humans are less confident than the AI-DSS, they show a higher propensity to change incorrect inferences to correct ones ($p <0.001$).
    \item Humans change from incorrect to correct inferences more when paired with AI-DSS that is well-calibrated.
    % ($p < 0.05$). Further, humans become more confident about their correct inferences when paired with a well calibrated AI-DSS ($p < 0.001$).
    % ($t(54) = 1.45, p < 0.05$)
    % ($ t(104) = 3.53, p < 0.001$).
    \item The MCS-based joint human–AI inference achieved higher accuracy than both human-initiated inference after observing AI-DSS ($p < 0.001$) and inference based on a Thompson sampling bandit ($p < 0.001$).
     % ($t(4828) = 2.02, p < 0.001$)
    \item MCS-based joint human-AI inference accuracy with a well-calibrated AI-DSS was significantly higher than that achieved with a poorly-calibrated AI-DSS ($p < 0.001$).
    % ($t(4828) = 18.04, p < 0.001$).
    
\end{enumerate}

% \begin{enumerate}
%     \item Misalignment between human and AI confidence significantly impairs joint decision-making accuracy ($p < 0.001$). 
%     \item MCS-based joint decision-making in the human-AI peer setting yields higher accuracy as compared to the AI-assisted setting with the human making the final decision ($p < 0.001$). Further, AI-DSS confidence calibration significantly affects the joint decision-making accuracy ($p<0.01$). 
%     \item AI-DSS significantly influences how humans change their decisions as well as their confidence based on the AI-DSS decision.
% \end{enumerate}

% Human decision-making, however, often relies on intuition, which can be susceptible to biases, emotions, and incomplete information \cite{casper2003human,norton2017analysis}. 

% \begin{figure}[t!]
%     \centering
%     \includegraphics[width=0.9\linewidth]{fig/short-meta-2hbt1.png}
%     \caption{\textbf{Joint Decision-Making Framework:} Human and AI-DSS shared choices and associated confidence levels for making joint decisions.}  
%     % \raunak{needs more description. Further, is something shared at both levels or only confidence?}}
%     % The diagram shows how confidence levels relate to metacognition and decision-making systems, with humans and AI exchanging confidence levels to make joint decisions. The top-level meta-decision determines whose decision becomes the final team choice. \an{remove metacognition system, monitor and control}}
%     \label{fig:joint-decision-via-metacog-1}
% \end{figure}
% \input{sections/background}
\section{Related Work}

Work in human inference with AI-DSS has been categorized based on the direction and presence of advice: AI advising the human, human advising the AI, and joint inference~\cite{zahedi2021human}.
In the former two cases, one of the two agents (either AI or human) makes the final inference and the other agent takes on the role of an advisor.
The final decision is made based on some representation of the capability of the AI system.
Confidence (equivalently uncertainty estimate) has been investigated in the literature as a measure of the capability of the AI system~\cite{jalaian2019uncertain,lai2023survey}.
Moreover, confidence calibration plays an important role as well-calibrated confidence can accurately reflect the actual likelihood of correctness of the AI~\cite{bansal2019updates}.

In the case where AI advises the human, confidence estimates associated with the AI-DSS recommendation as well as self-confidence of the human have both been shown to influence the propensity of humans to accept/reject the AI-DSS inference~\cite{ma2024you,benz2023humanaligned}.
Humans are shown a probabilistic algorithmic prediction to incorporate into their decision in domains such as energy~\cite{shin2021ai}, medicine~\cite{kiani2020impact}, and classification tasks on images and fake news~\cite{cabrera2023improving}. 
For example,~\citet{marusich2024using} studied tasks where humans had to make judgements on the Census, German Credit and Student Performance datasets given recommendations and confidence scores from classification models.
Similarly,~\citet{benz2023humanaligned} studied binary prediction tasks such as choosing between two cities given an image, choosing the period of an art painting given two choices and choosing whether a piece of text contains sarcasm.

Confidence-based approaches have also been investigated in the case where the human plays an advising role~\cite{kerrigan2021combining,gupta2023take}.
For example,~\citet{madras2018predict} studied recidivism risk prediction and patient's comorbidity indicator prediction tasks where the classification system's confidence was used to decide to defer to a human expert.
Similarly, ~\citet{mozannar2020consistent} studied detecting hate speech and offensive text as well as detecting the presence of medical conditions from radiology reports using confidence-based deferral to human experts.

However, instead of fixing the role of inference maker and advise giver to either the human or the AI, a third line of work has investigated joint inference where either the human or the AI-DSS's inference is selected.
This has drawn inspiration from the work on joint human-human decision making where neither human takes on an advisory role and a joint decision-making framework is established instead.
Confidence-based joint decision making has also shown promise in this case.
In joint visual perception tasks such as locating a target object between two images placed side-by-side for a short duration (oddball target detection), the maximum confidence slating approach has yielded higher team accuracy than either individual human alone~\cite{bahrami2010optimally,koriat2012two}.
Using MCS, enhanced team performance outperforming both individual humans has been shown in varied joint human-human inference tasks such as threat detection~\cite{2hbt1_threatdetection}, detecting fake news~\cite{2hbt1_fakenewsdetection}, deciding rank ordering between items on a survival situation task~\cite{2hbt1_nasasurvival}, breast and skin cancer diagnosis~\cite{2hbt1_cancerdiagnosis} as well as robot missions~\cite{nguyen2025group}.

Since many real-world robotics operations involve multiple human operators, recent work has investigated the MCS approach on human-human joint inference for a robotics task~\cite{nguyen2025group}.
Crucially, the confidence calibration of individual decision makers influences the extent of the benefit. 
Of particular interest is the characterisation of human-AI teams in terms of the locus of control~\cite{jiang2015mixed} for robotic teleoperation tasks.
While in human-initiative control, the human operator is solely responsible for the final team decision, in mixed-initiative control, the eventual decision can be made by either the human or the AI-DSS~\cite{chiou2021mixed}.
While mixed-initiative control has found to yield benefits in terms of user preferences and performance, it has also introduced the problem of conflict for control where the human and the AI-DSS disagree on the decision to be made~\cite{ramesh2023experimental}. 
This motivates our study on choosing between the human and AI-DSS inferences using MCS.

\section{Methodology}

% This section details the robotic settings used in our study; the experiment procedure; the development of AI-assisted intelligent decision-support systems, with a specific focus on the confidence-sharing framework explored in this paper, we also presents the meta-decision approach for resolving human-AI disagreements using the Maximum Confidence Slating (MCS) algorithm, along with the evaluation metrics.

\subsection{Problem Statement}
Consider two agents, a human and an AI-DSS who, given an input $x$, produce calibrated probabilistic predictions $P_{human}(x), P_{AI}(x) \in [0,1]$.
Calibrated means that among inputs for which the agent predicts a positive label with probability $p$, the true proportion of positive labels is in fact $p$.
Each individual agent, i.e., the human and the AI-DSS can achieve some level of accuracy on their own.
For the agent $M$, given their calibrated prediction $P_M(x)$, the optimal binary classification to maximise accuracy is obtained using a threshold rule: predict $0$ if $P_M(x) < 0.5$ and $1$ if $P_M(x) > 0.5$.

A joint inference strategy is a way to combine the calibrated predictions $P_{human}(x), P_{AI}(x)$ into a binary classification.
The MCS heuristic uses the following collaboration strategy: accept the classification of the most confident agent, i.e., the agent whose probabilistic prediction is furthest from $0.5$.
This collaboration strategy also achieves an accuracy, which can be compared to the accuracy of either individual agent.
Human-AI complementarity is achieved when the joint inference strategy produces binary classifications that are at least as accurate than either individual agent~\cite{peng2025no}.

On a representative joint inference task (detailed next), the human-AI-DSS team makes a joint inference using the MCS approach. We seek to find whether the team inference accuracy is higher than: 1) The human making the inference alone, and 2) The human making the inference in the AI-assisted setting.
Further, does well-calibrated AI-DSS lead to better team inference accuracy?

\subsection{Environment and Task}
\begin{figure}[]
    \centering
    \includegraphics[width=0.9\linewidth]{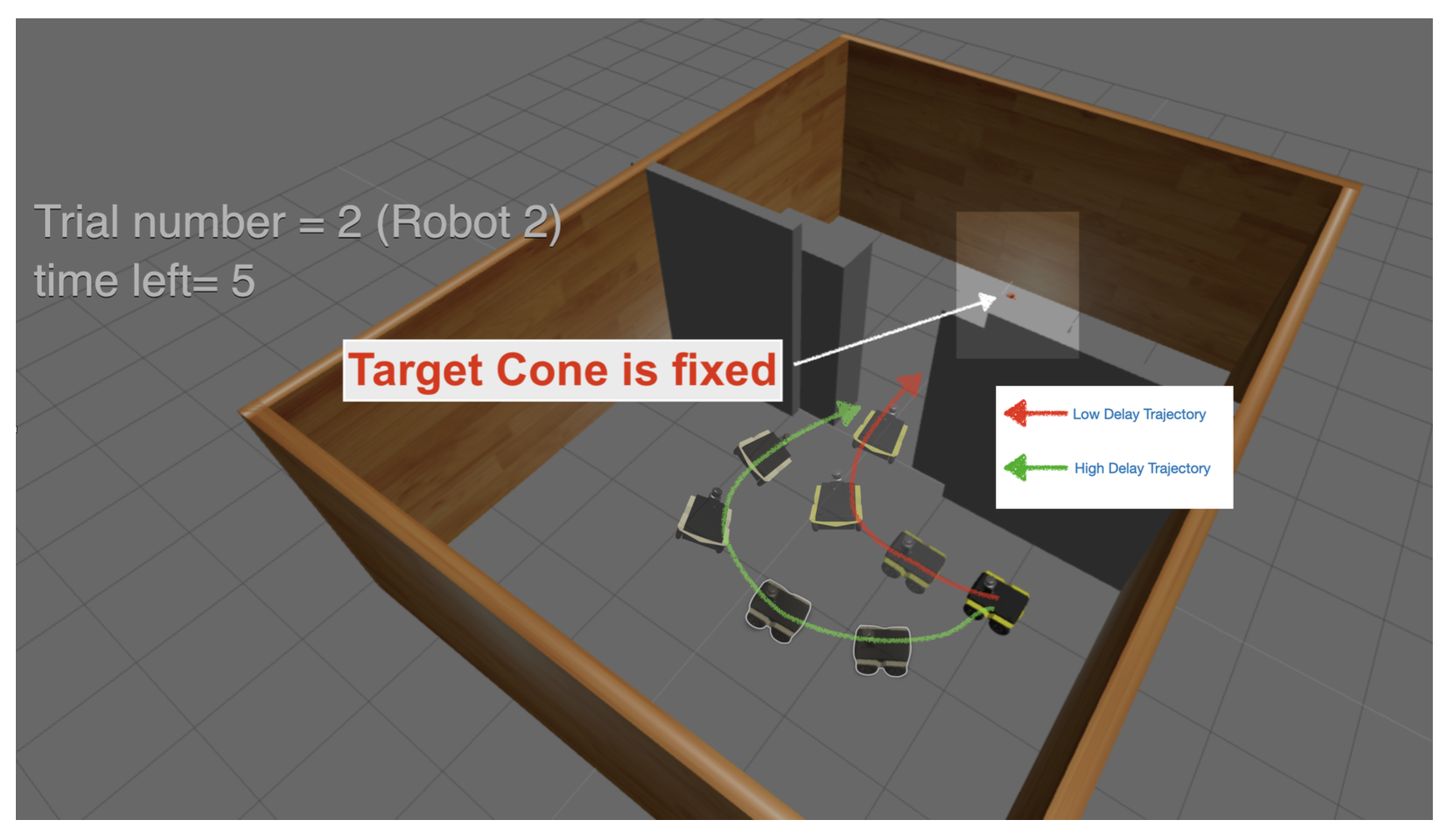}
    \caption{\textbf{Experiment Setup:} Online robot navigation simulator. Red and green curves show trajectories driven by a participant controlling the Jackal robot under low and high delay conditions, respectively.}
    \label{fig:robot-driving-task}
\end{figure}

We consider the following joint human-AI inference task. Human participants complete a set of $N_{trials}$ binary robot selection tasks in varying environments with the aid of an AI-DSS. 
In each trial, human participants teleoperate two identical robots. 
The only difference between the two robots is the delay in the controller, which is an unobserved latent variable.
They control the robot from a start location to the goal through a narrow doorway. 
At the end of each trial, participants must select the robot that is more manoeuvrable, i.e., the one with the lower controller delay (see Fig~\ref{fig:robot-driving-task}).

At environment reset 
% after $t$ ($1 \leq t \leq$ $n_{trials}$) trials, 
the teleoperated robot starts in one of six possible initial positions, and the doorway has four different gap configurations, resulting in diversified 24 environment settings. 
The goal position, to which the human must teleoperate the robot within the time limit, remains fixed and marked by a traffic cone. 
Two example trajectories of the robot teleoperation task are shown in Fig.~\ref{fig:robot-driving-task}.

The delay characteristics of each robot follows a staircase procedure. 
In this procedure, the task difficulty is adjusted adaptively: after two consecutive successful inferences, i.e., selecting the robot with lesser delay, the delay difference between the two robots is decreased by 20 ms, making the behaviour difference between the two robots harder to distinguish. 
Conversely, after an incorrect choice, the delay is increased by $\Delta_d = 20 ms$, making the task easier. 
This adaptive adjustment is standard in human behavioral studies on confidence and ensures that the accuracy achieved by any participant is around 70\%, preventing the task from being too easy (resulting in ceiling performance) or too difficult (leading to performance close to random)~\cite{fleming2014measure,rouault2018human}. 
The control delay differential between the robots starts at 35 ms and progresses through five levels of task difficulty: 20 ms, 40 ms, 60 ms, 80 ms, and 100 ms differences in the robots' control delays. On average, it takes a participant 60 minutes to complete 100 trials. Gazebo was used as the simulation environment, with the Jackal robot as the model.

\subsection{AI Decision Support System (AI-DSS)}
% \raunak{Task difficulty role in this subsection as well?}\an{Task difficulty does not do much to AI learning stratergy. Task difficulty was used later in our bandits works, not this one}
\begin{figure}[h!]
    \centering
    \includegraphics[width=0.95\linewidth]{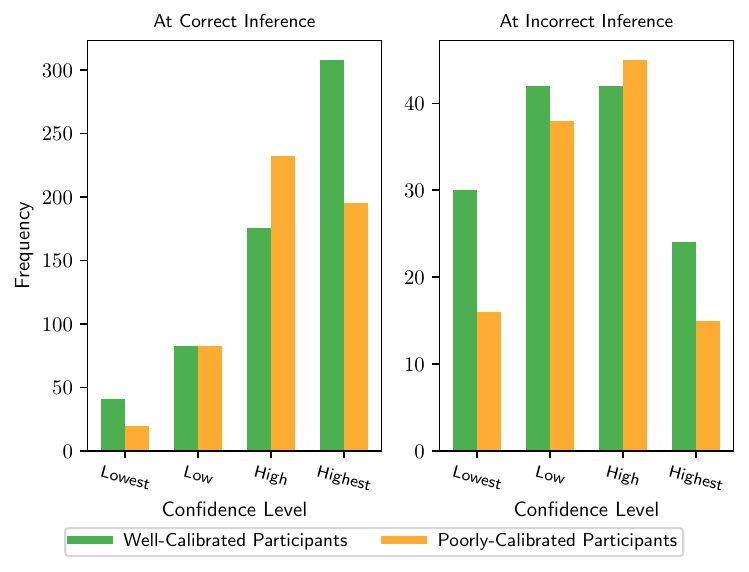}
    \caption{Histograms of human confidence levels associated with correct and incorrect inferences and separated by confidence calibration of the participants. The delay difference between the two robots was 60 ms in this case.}
    % \an{need to discuss if to show both 40ms and 60ms, or just 40ms - Done}}
    \label{fig:conf-hist}
\end{figure}
\begin{figure*}[t!]
    \centering
    \includegraphics[width=0.8\linewidth]{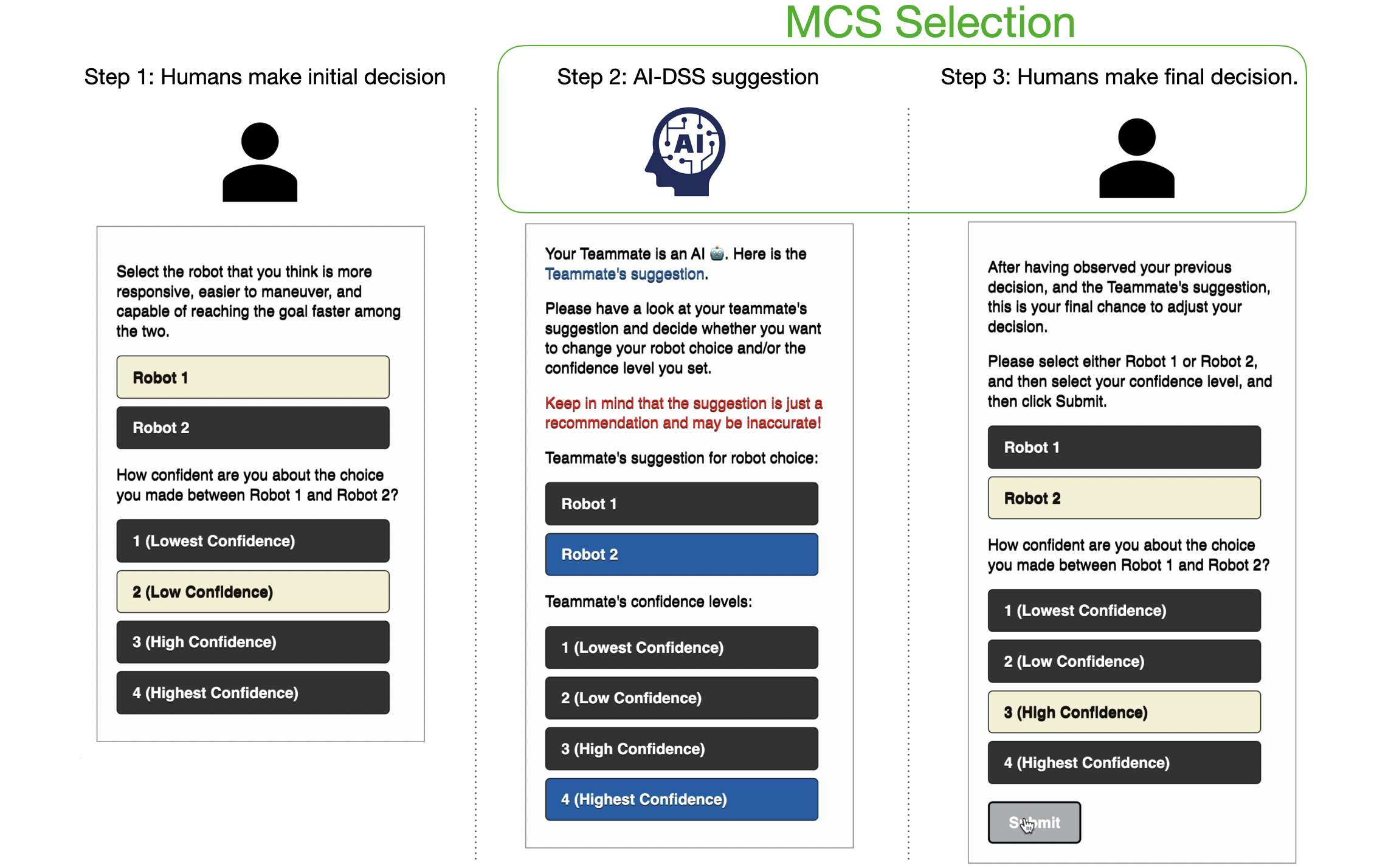}
    \caption{\textbf{Study Interface:} After teleoperating the two robots, participants make an initial inference and provide associated confidence (left panel). Then, the AI-DSS provides its inference and confidence (middle panel). If the participant decides to change their initial inference, they provide an updated inference and associated confidence (right panel). The joint human-AI inference is then decided using MCS (top right).}
    % }
    % This final choice is considered the human's self meta-decision.}
    % \an{human final decision is considered as one kind of meta-decision.}}
    \label{fig:user-interface}
\end{figure*}

To construct the AI-DSS, we collected data from 100 participants who performed the same task without the presence of an AI-DSS.
For each set of robot delay characteristics, which determine the difficulty of the task 
% (e.g. Detecting a 40ms control delay different (Level 1) is harder than 60ms control delay different (Level 2), 
we separated the data into two sets of trials where the humans made correct and incorrect inferences, respectively. 
Fig.~\ref{fig:conf-hist} shows an example of the discrete distribution over the human participants' confidence level associated with them making correct and incorrect inferences for a delay difference of 60 ms. 
% The distribution of confidence changed with difficulty level.
During the human-AI-DSS joint inference tasks, the AI-DSS samples its confidence from these discrete distributions conditioned on the task difficulty level and accordingly provides an associated confidence score.

To study the impact of confidence calibration, we constructed both a well-calibrated and a poorly-calibrated AI-DSS.
We use AUROC2 as a measure of confidence calibration\footnote{Further details on computing the AUROC2 values are provided in the Supplementary Material.} \cite{steyvers2025metacognition,steyvers2025large,fleming2014measure}. AUROC2 scores range from 0 to 1.0. An AUROC2 of 0.85 means there's an 85\% chance that a person gives higher confidence to correct answers than to incorrect ones \cite{sherman2024knowing}. Unlike some correlation measures, AUROC2 is a robust and bias-free metric - it remains largely unaffected by whether someone tends to give generally high or low confidence ratings \cite{katyal2024future}.

In our study, which involves teleoperating robots and inferring control differences from partial information, human participants showed a wide range of AUROC2 scores - from 0.45 to 0.85, with an average of 0.65. Based on this, we divided the dataset into two groups for AI-DSS learning: the well-calibrated AI-DSS samples from the data of humans with well-calibrated confidence (AUROC2 $\geq$ 0.65). The poorly calibrated AI-DSS sample from the data of participants with poorly-calibrated confidence (AUROC2 $\leq$ 0.55).
% \textcolor{blue}{AUROC2 number means things, cite paper on the range -Done}
% \begin{equation*}
%     \texttt{AUROC2} = \sum_{i=1}^{n-1} (\tt FPR_{i+1} - FPR_i) \cdot \frac{TPR_{i+1} + TPR_i}{2} \text{,}
% \end{equation*}
% where $n$ is the number of confidence levels on the Likert scale. $\texttt{TPR}$ and $\texttt{FPR}$ are the True Positive Rate and False Positive Rate of the inference at each confidence level, represented by $i$.
We fixed the AI-DSS accuracy at 70\% to simulate a realistic AI-based decision support system that may be imperfect.
% Given its current knowledge of human confidence and inferencing accuracy, the AI recommends choices to human users, along with its confidence level for each separate tele-operation trial.

% \textbf{Joint Inference using MCS:}
% % On the decision-making scenarios where the human participant and AI agent held opposing choice of robot. To maximize the best outcome in these cases, we employed the Maximum Confidence Slating (MCS) algorithm, which is informed by prior research in the psychology domain as the meta-decision maker~\cite{massoni2017optimal, bang2017confidence, rouault2018human}. Specifically, the MCS approach prioritizes the decision of individuals in which the presented the \textit{higher confidence}~\cite{bahrami2010optimally,koriat2012two}. 
% On the inferencing scenarios where the human participant and AI held opposing choices of robot and had difference on confidence levels, we needed to determine whose inference should prevail. We employed the Maximum Confidence Slating (MCS) algorithm as the joint inference. This approach selects the choice in which the participant has the \textit{higher confidence} levels.
% % , which is informed by prior research 
% % in the psychology domain ~\cite{massoni2017optimal, bang2017confidence, rouault2018human}. 

\subsection{Experimental Procedure}

We conducted an online study, approved by the University Research Ethics Committee for data collection.
Responses from 100 participants in total were collected via the Prolific online platform\footnote{https://www.prolific.com/}. 
Each participant performed 100 trials.

First, the participants received a detailed study overview, including a video demonstration and example trial runs. 
They then completed 5 practice trials to familiarise themselves with controlling the robot using the keyboard controller. 
Next, they performed the 100 trials. 
In each trial, they teleoperated a pair of robots.
After each trial, they made an initial inference, i.e., they selected the robot they deemed having the lower control delay, and provide their confidence scores.
The AI-DSS, designed as described in section 3.3, then provided the human participant with its inference along with its confidence. 
The participants were then asked to make a final inference after having observed the AI-DSS inference and associated confidence.
Both the human participant and the AI-DSS provided their confidence using a four-point Likert scale, where 1 indicated the lowest confidence and 4 the highest.
The experimental flow and study interface is shown in Fig.~\ref{fig:user-interface}.
\textcolor{blue}{}

After data collection, we excluded participants who were either unable to perform the task or were inattentive, based on the following criteria: (i) those with an accuracy of less than 65\%, and (ii) those who gave the same confidence rating response for over 95 of 100 trials. 
Inattentive participants could achieve above-chance performance by guessing, so setting the threshold at 65\% excluded them while retaining engaged participants. This is standard procedure in studies on human self-confidence~\cite{rouault2018human}.
Out of the 100 participants, 80 met the above criteria.

\section{Results}

We report our findings on: a) the influence of AI-DSS performance and confidence on human inference and confidence, and b) joint inference accuracy when MCS is used to determine the final inference by selecting between the human and AI-DSS inference according to higher confidence.
% \begin{figure}[]
%     \centering
%     \includegraphics[width=0.7\linewidth]{fig/crop-rate-decision-changed-2.png}
%     \caption{Higher confidence scores from the AI-DSS result in a higher number of inferences changed by humans.}
%     % \an{make new negative change and positive changes according to AI confidence level - Updating: Figure 5}}
%     \label{fig:C-1}
% \end{figure}

%**************IMPACT ON HUMAN DECISION MAKING********************
\subsection{Human Inference Dynamics}

% We first investigate the frequency with which humans change their initial inference based on the AI-DSS confidence level.
% In other words, does the confidence score associated with the AI-DSS inference in each trial impact humans' propensity to change their original inference of which robot they thought had a lesser delay?

% Fig.~\ref{fig:C-1} shows that humans exhibited the strongest tendency to revise their initial inferences when presented with AI-DSS recommendations accompanied by the \textit{highest confidence levels} ($14.9 \%$ of trials). 
% On the other hand, when the AI-DSS confidence level was low, the percentage of trials where the human participants changed their inference dropped to $3.8\%$.
% \raunak{How many trials in each category of AI confidence?}

We first analyse the quality of human inference revision according to the AI-DSS confidence calibration quality. According to each poorly and well confidence-calibrated AI-DSS, did the humans change their inferences to correct ones or did they end up changing from an initial correct inferences to an incorrect one?
We defined \textit{positive changes} as instances where humans initially made incorrect inferences but, after observing the AI-DSS inference and its confidence, revised their original choice to a correct inference.
Conversely, \textit{negative changes} occurred when humans initially made correct inferences but changed to incorrect inferences after observing the AI-DSS inference and its confidence. 
The number of each type of change was counted against the total number of changes to calculate the percentage of each type.
% We analysed how the percentage of \textit{positive} and \textit{negative} inferences changes were influenced by: 1) the difference in confidence level between humans and AI-DSS, and 2) the confidence calibration of the AI-DSS.
\begin{figure}{}
    \centering
    \includegraphics[width=0.9\linewidth]{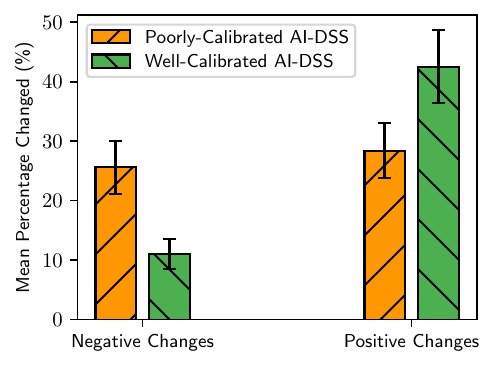}
    \caption{Dynamics of human inferences: AI-DSS confidence-calibration influences whether humans change from incorrect to correct inferences.}
    \label{fig:C-2}
\end{figure}
Fig.~\ref{fig:C-2} shows the mean percentage of positive and negative inference changes according to the confidence calibration of the AI-DSS. 
The analysis reveals that well-calibrated AI-DSS systems facilitated a significantly higher percentage of positive inference changes ($42.53\% \pm 6.19\%$) vs poorly-calibrated AI-DSS ($28.41\% \pm 4.64\%$), whereas poorly calibrated AI systems induced negative inference changes at a significantly higher percentage ($25.58\% \pm 4.46\%$) vs well-calibrated AI-DSS ($10.95\% \pm 2.50\%$). 
\begin{figure}[h]
    \centering
    \includegraphics[width=\linewidth]{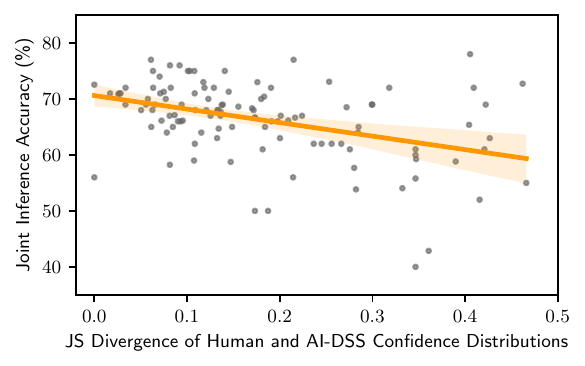}
        \caption{Misalignment in human and AI confidence ratings leads to degradation in accuracy of human-initiative joint inference.}
        % : The greater divergence in confidence alignment between humans and AI systems consistently predicted lower team joint decision accuracy ($r = 0.183; p<0.001$).}
        % \an{crop fig}}
    % ($p < 0.001$)
    \label{fig:A-1}
\end{figure}

We also analysed the relationship between confidence alignment and human-initiative inference performance. The Jensen-Shannon divergence (JS) was used to quantify distributional differences between human and AI-DSS confidence levels. Higher JS divergence values indicate greater dissimilarity and, consequently, increased misalignment between human and AI-DSS confidence.
% This analysis employed established distributional distance metrics, including Jensen-Shannon Divergence, which is well-suited for quantifying differences between probability distributions.
For each of the 80 participants, we calculated JS divergence of human confidence's distribution from AI-DSS confidence distribution across all task difficulty levels (defined by the delay differential between the two robot controllers as mentioned previously) and then averaged the value for the five levels of difficulty to compute the averaged JS divergence. This average distributional differences were then plotted against joint human-initiative inference accuracy, with the aggregate data fitted using linear regression Fig.~\ref{fig:A-1} revealed statistically significant relationships ($r^2=0.183, p<0.001$) demonstrating that greater divergence in confidence alignment between humans and AI-DSS predicts lower joint human-initiative inference accuracy.
% when paired with well calibrated AI, increased human confidence reliably predicted higher decision accuracy. In contrast, poorly calibrated AI frequently induced higher confidence while simultaneously decreasing decision accuracy.

% \textcolor{blue}{add extra results into Sup Mat}

%****************JOINT DECISION PERFORMANCE*****************
\subsection{Joint Inference Performance}

% \an{Since human-initiative isn't the optimal way to go. How can we leverage MCS}
% At testing aaply the MCS method in making the final decision where team member disagreeing on choice and confidence, we found that humans making the final decision is often ineffective.
% Fig.~\ref{fig:B-1} is the boxplot of accuracy across 100 trials for all participant, having the final joint decision decided by human, AI (both good and bad) and the MCS algorithm for total 5 levels of task difficulty. It can be seen that such a simple, easily accessible confidence metrics-based method, such as MCS, delivers better joint decision-making outcomes the either of individuals in the dyad ($p<0.01$). The affect was more prominent in the trial with task difficulty 3-5. Therefore, we provde the insight where MCS best aplpied the effectiveness of confidence-sharing methods like MCS depends on scenario complexity or specifically task difficulty.
% It's been observed that relying solely on human for final decisions often yields suboptimal results. Human-initiative as joint decision maker is not the most optimal method. 
\begin{figure}[h!]
    \centering
    \includegraphics[width=\linewidth]{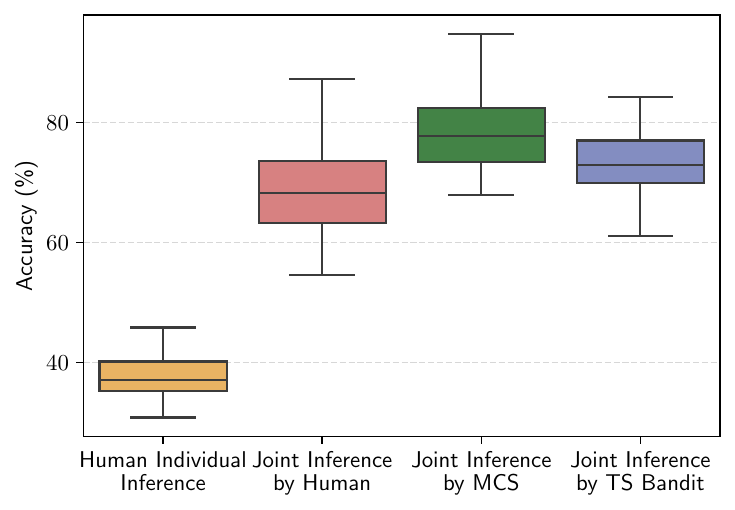}
    \caption{Joint human-AI inference accuracy is highest with MCS, outperforming purely human individual inference, joint inference dictated by human, and fusion approaches using Thompson Sampling (TS) bandits.}
    % Joint decision-making human-AI team achieved higher accuracy than human decision-making alone ($p< 0.001$). Moreover, the MCS combining human and AI decisions outperformed joint decisions decided human ($p<0.001$).}
    \label{fig:B-1}
\end{figure}
% At testing aaply the MCS method in making the final decision where team member disagreeing on choice and confidence, we found that humans making the final decision is often ineffective.
We present our benchmark results on joint human-AI-DSS inference performance.
We compare the performance of Maximum Confidence Slating (MCS) against human-initiative inference with AI-DSS support as well as against individual human inference without AI-DSS support.
The performance was measured using accuracy, i.e., the proportion of correct inferences made relative to the total number of inference trials.

Fig.~\ref{fig:B-1} presents results from aggregated data of 80 human participants interacting with both well- and poorly-calibrated AI-DSS. These results focus on cases where the human and AI-DSS provided opposing inferences that required resolution to produce a final joint team decision (e.g. Human chose Robot 1, while AI-DSS suggested Robot 2).
`\textit{Joint Inference by Human}' represents cases the final inference is made by the human participant (i.e., human-initiative). `\textit{Joint Inference by MCS}' refers to cases where the final inference is determined by the MCS. `\textit{Joint Inference by TS Bandits}' refers to the case where a Thompson sampling bandit algorithm chooses the final inference, selecting between the human’s or AI-DSS’s choice. We observe that human participants improved their inference accuracy after receiving AI-DSS recommendations compared to their initial individual decisions. However the MCS-based joint inference significantly outperformed both human-initiated joint inference ($t(4828) = 8.52$, $p < 0.001$) and Thompson sampling bandit approach ($t(4828) = 4.74$, $p < 0.001$).
All t-tests in our results are 2-sided.

This finding highlights that relying solely on human inference is suboptimal. Instead, MCS provides a simple yet robust method for enhancing team inference by leveraging a readily available metric - confidence.

\begin{figure}[h]
    \centering
    \includegraphics[width=0.9\linewidth]{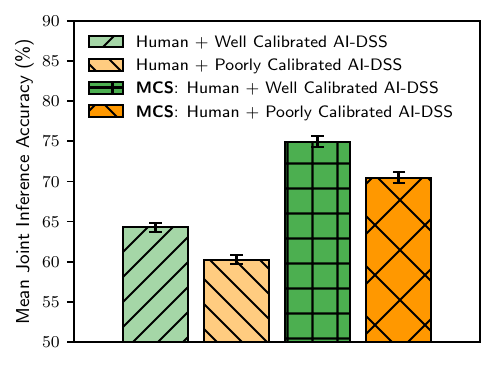}
    \caption{Influence of varying AI type on team inference accuracy.}
    % Well-calibrated AI consistently enhances joint decision-making accuracy in both easy and hard tasks when humans act as meta-decision makers. Poorly calibrated AI leads to weaker performance, while MCS ensures  as a meta-decision-making approach. 
    % \an{change y-axis title, typo x-axis}}
    \label{fig:B-2}
\end{figure}
We further analyzed the effect of AI-DSS confidence calibration on the performance of MCS-based joint human-AI inference. Participants were paired with either a well-calibrated AI-DSS (AUROC2 $\geq$ 0.65; 42 participants) or a poorly calibrated AI-DSS (AUROC2 $\leq$ 0.55; 38 participants).

Fig.~\ref{fig:B-2} shows joint inference accuracy for both the MCS-based and human-initiated approaches, separated by the AI-DSS calibration condition. When paired with a well-calibrated AI-DSS, the MCS-based joint inference achieved an accuracy of ($64.28\% \pm 0.59\%$), significantly outperforming the human-initiated joint inference, which reached ($60.24\% \pm 0.49\%$). Moreover, MCS performance was also sensitive to the calibration quality of the AI-DSS. MCS-based joint inference with a well-calibrated AI-DSS yielded ($74.95\%\pm0.68\%$) accuracy, which was significantly higher than the ($70.47\% \pm 0.65\%$) achieved when paired with a poorly calibrated AI-DSS. These findings demonstrate two key insights:
\begin{enumerate}
    \item MCS consistently outperforms human-initiated resolution, confirming the value of confidence-based joint inferencing.
    \item The effectiveness of MCS is amplified when paired with a well-calibrated AI-DSS, highlighting the critical role of AI-DSS's confidence calibration in human–AI team performance.
\end{enumerate}

\begin{figure}[h]
\centering
\includegraphics[width=0.9\linewidth]{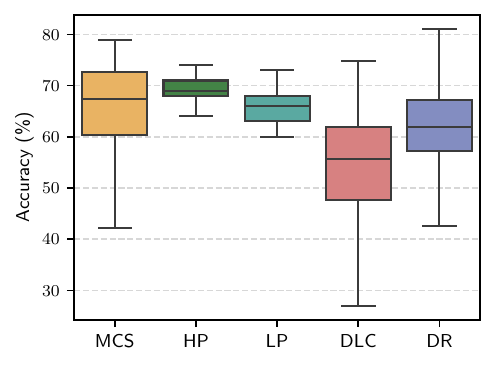}
\caption{Inference accuracy obtained by pairing human participants with the poorly confidence-calibrated AI-DSS.}
\label{fig:B-10}
\end{figure}

% We then analyse the effect of having well confidence calibrated AI and poorly confidence calibrated AI in assitting human in making the decision, seperately for hard task diffculty level (Level 1 \& 2 ) and easy task difficulty level (3, 4 \& 5). The results is depicted in Fig.~\ref{fig:B-2}, with accuracy gained boxploted for 2 type of AI agents accross hard and easy task scenario.

% It can be observed that well confidence calibrated AI agent provide significant accuracy gain in the joint decision making process, whereas poorly confidence calibration AI agent was able to help on easy tasks, but can effect bad toward hard task. Sometimes hamper human decision accuracies, making negative accuracy gain. 

% Finally we specifically looking into the performance at want team inference performance on the basis of confidence calibration of both the human and the AI-DSS.
% We divided the group of human participants into two categories: well-calibrated (AUROC2 performance $\geq$ 0.65) and poorly-calibrated (AUROC2 performance $\leq$ 0.55).
% Accordingly, we identified trials belonging to three categories of human-AI-DSS pairs according to confidence calibration: 1) Both well calibrated, 2) Well-calibrated human paired with poorly-calibrated AI-DSS, and 3) Poorly-calibrated human paired with well-calibrated AI-DSS, and assessed the performance of MCS-based joint inference.
To better understand why MCS did not perform well when paired with poorly calibrated AI-DSS, we identified all trials where the AI-DSS was poorly calibrated. We created a virual pairing of the poorly-calibrated AI-DSS with data from all human participants who operated under the same robot control delay differential. 
This virtual pairing allowed us to compute joint inference accuracy across consistent task difficulty levels.

Within each virtual human-AI dyad, we investigated three different joint inference strategies: MCS, Dummy Low Confidence (DLC), which selects the low-confidence inference, and Dummy Random (DR), which randomly selects between the human and AI-DSS inferences. 
We also included two baselines: High-Performing (HP), the individual in the dyad with a higher percentage of correct responses across trials, and Low-Performing (LP), the less accurate individual. Fig.~\ref{fig:B-10} presents the accuracy of these joint inference strategies. 
HP slightly outperformed MCS ($t(1716) = -1.15, p < 0.001$). 
This outcome reflects an important limitation of MCS. 
When MCS is applied with poorly calibrated AI-DSS, it selects the AI-DSS inference due to its inappropriately high confidence, even when the AI-DSS is incorrect. 
Thus, MCS overrides the human's correct judgment. 
This result underscores the importance of confidence calibration for the effective functioning of confidence-based methods such as MCS.

\section{Discussion}
As our first research question, we asked how AI-DSS inference and confidence impacted human inference and confidence.
% \textcolor{blue}{remove sentence without associated figs} 
Our results give interesting insights into human making joint inference with an AI-DSS. 
% Firstly, humans changed inferences more when the AI-DSS associated its inference with a higher confidence score (Fig.~\ref{fig:C-1}). 
% Second, humans showed a higher propensity to change to correct inferences from incorrect ones when the AI-DSS was more confident in its inference (Fig.~\ref{fig:C-8}).
First, the confidence calibration of the AI-DSS impacted the nature of the human inference change (Fig.~\ref{fig:C-2}).
A well-calibrated AI-DSS led to humans making more positive changes, i.e., changing from an initially incorrect inference to a correct one. Second, misalignment in confidence between the human and the AI-DSS led to reduced inference accuracy when the human made the final inference after having observed the AI-DSS inference (Fig.~\ref{fig:A-1})
% Further, a well-calibrated AI-DSS encouraged humans in their correct inferences, evidenced by a greater increase in human confidence, while poorly calibrated AI-DSS induces humans to become more confident in their incorrect choices.  
This finding underscore that while the provision of confidence scores alongside AI-DSS inferences can be beneficial, the degree to which this information improves joint human-AI inference is critically dependent on the underlying confidence calibration of AI-DSS.

As our second research question, we asked how joint human-AI-DSS inference performed relative to AI-assisted inference.
First, joint human-AI inference accuracy was highest with MCS, outperforming both AI-assisted inference, purely human inference and the Thompson sampling bandit (Fig.~\ref{fig:B-1}).
Second, MCS with well-calibrated AI-DSS yielded better joint inference outcomes as compared to both MCS with poorly-calibrated AI-DSS and human final inference with poorly-calibrated AI-DSS (Fig.~\ref{fig:B-2}). Third, we identified the limitation (poorly-calibrated AI-DSS) that prevents the heuristic method such as MCS from achieving fully complementary performance (Fig.~\ref{fig:B-10}).
% Third, misalignment in confidence between the human and the AI-DSS (as measured by JS divergence) led to reduced inference accuracy when the human made the final inference after having observed the AI-DSS inference (Fig.~\ref{fig:A-1}).
% Finally, we showed the efficacy of MCS for joint inference even in less-than-ideal pairings where poorly-calibrated humans were paired with poorly-calibrated AI-DSS (Fig.~\ref{fig:B-10}).
These findings show the potential and dependency of MCS-based joint human-AI inference towards achieving complementarity.

While the proposed approach and designed user study provides a valuable first step for confidence-guided human-AI joint inference in robot missions, there are several limitations. 
First, our study assumes that both the AI-DSS and the human recieve the same information. However, in real-world situations, they may have different sources and amounts of information leading to the dependence of confidence on context.
Second, the AI-DSS is constructed based on separately collected human data. In future work, we will investigate learning-based and pretrained AI-DSS.
Finally, our study considers a relatively simple task. This was not only to enable a behavioral study with a large number of participants-ensuring statistically significant results-but also because we carefully designed the task to isolate the variable under investigation in a controlled manner. The simplicity of the task allows us to draw clear insights into the mechanisms of joint human-AI inference. That said, future work will extend this investigation to more complex, multi-step sequential decision-making tasks, such as path safety discrimination or signal source localisation, where joint human-AI inferences may be used as part of a longer sequence of actions.

\section{Conclusion}
Our study makes a valuable addition to the literature on joint human-AI inference using confidence-based approaches.
First, we adapt the MCS-based approach of selecting the inference made with higher confidence to the setting of joint human-AI inference.
Second, we show that this approach improves team performance in a representative robotics task that involves inferring a latent variable by taking actions and observing the resulting temporal process.
Our results show that trialwise confidence is a better predictor for accuracy than overall performance.
This finding can be applied across a variety of novel applications of human-AI joint decision-making where there are either no accuracy data yet or accuracy data varies strongly with task characteristics.
We hope that our study lays the groundwork for further research on approaches to improve human-AI collaborative decision-making resulting in safer, more efficient, and more reliable human-supervised robotic operations.

\bibliography{aaai-26}
% \balance
\newpage

% \begin{figure*}[htbp]
%   \centering
%   \includegraphics[width=\linewidth]{fig/aaai_check_list1.pdf}
%   % \caption{Your caption here}
%   \label{fig:yourlabel}
% \end{figure*}

% \begin{figure*}[htbp]
%   \centering
%   \includegraphics[width=\linewidth]{fig/aaai_check_list2.pdf}
%   % \caption{Your caption here}
%   \label{fig:yourlabel}
% \end{figure*}
\end{document}